\documentclass[aps,prl,twocolumn,superscriptaddress]{revtex4-2}

\usepackage{graphicx}
\usepackage{epsfig}
\usepackage{dcolumn}
\usepackage{bm}
\usepackage{overpic}
\usepackage{color}
\usepackage{multirow}
\usepackage{subfigure}
\usepackage{lineno}
\usepackage{amsmath}
\usepackage{url}
\usepackage{ulem}

\usepackage{xcolor}

\usepackage{hyperref}
\hypersetup{
colorlinks=true,
linkcolor=blue,
filecolor=blue,
urlcolor=blue,
citecolor=blue,
pdftitle={LaTeX Example},
pdfpagemode=FullScreen,
}

\newcommand{\BESIIIorcid}[1]{\href{https://orcid.org/#1}{\hspace*{0.1em}\raisebox{-0.45ex}{\includegraphics[width=1em]{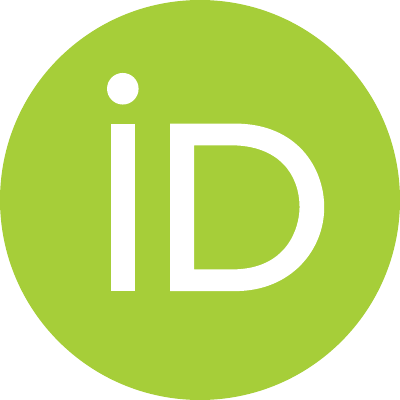}}}}

\begin{document}


\title{Proof of principle for nucleon polarization measurement at BESIII}

\author{
M.~Ablikim$^{1}$\BESIIIorcid{0000-0002-3935-619X},
M.~N.~Achasov$^{4,c}$\BESIIIorcid{0000-0002-9400-8622},
P.~Adlarson$^{81}$\BESIIIorcid{0000-0001-6280-3851},
X.~C.~Ai$^{87}$\BESIIIorcid{0000-0003-3856-2415},
C.~S.~Akondi$^{31A,31B}$\BESIIIorcid{0000-0001-6303-5217},
R.~Aliberti$^{39}$\BESIIIorcid{0000-0003-3500-4012},
A.~Amoroso$^{80A,80C}$\BESIIIorcid{0000-0002-3095-8610},
Q.~An$^{77,64,\dagger}$,
Y.~H.~An$^{87}$\BESIIIorcid{0009-0008-3419-0849},
Y.~Bai$^{62}$\BESIIIorcid{0000-0001-6593-5665}
% ...
\\
\vspace{0.2cm}
(BESIII Collaboration)\\
\vspace{0.2cm} {\it
$^{1}$ Institute of High Energy Physics, Beijing 100049, People's Republic of China\\
% ...
$^{4}$ Budker Institute of Nuclear Physics SB RAS (BINP), Novosibirsk 630090, Russia\\
% ...
$^{31}$ INFN Sezione di Ferrara, (A)INFN Sezione di Ferrara, I-44122, Ferrara, Italy; (B)University of Ferrara, I-44122, Ferrara, Italy\\
% ...
$^{39}$ Johannes Gutenberg University of Mainz, Johann-Joachim-Becher-Weg 45, D-55099 Mainz, Germany\\
% ...
$^{62}$ Southeast University, Nanjing 211100, People's Republic of China\\
% ...
$^{64}$ State Key Laboratory of Particle Detection and Electronics, Beijing 100049, Hefei 230026, People's Republic of China\\
% ...
$^{77}$ University of Science and Technology of China, Hefei 230026, People's Republic of China\\
% ...
$^{80}$ University of Turin and INFN, (A)University of Turin, I-10125, Turin, Italy; (B)University of Eastern Piedmont, I-15121, Alessandria, Italy; (C)INFN, I-10125, Turin, Italy\\
$^{81}$ Uppsala University, Box 516, SE-75120 Uppsala, Sweden\\
% ...
$^{87}$ Zhengzhou University, Zhengzhou 450001, People's Republic of China\\

\vspace{0.2cm}
$^{\dagger}$ Deceased\\
% ...
$^{c}$ Also at the Novosibirsk State University, Novosibirsk, 630090, Russia\\
% ...
}
}

\date{\today}

\begin{abstract}

A novel technique for measuring the spin polarization of final-state
nucleons in a general-purpose spectrometer is validated. Using
$10.09\times10^{9}$ $J/\psi$ events at BESIII, the asymmetry of
polarized proton scattering on detector support material is measured,
and is consistent with the expected value. This proves that a
general-purpose spectrometer can be utilized as a large-acceptance
polarimeter, providing the spin polarization in addition to the
conventional four-momentum information of the final-state
particles. With this technique, physics capabilities are enhanced for
existing and future facilities in particle and nuclear physics.

\end{abstract}

\maketitle

\section{Introduction}

In nuclear and particle physics, large-acceptance general-purpose
spectrometers are widely used in collider experiments or fixed-target
experiments, such as BESIII~\cite{ABLIKIM2010345},
Belle-II~\cite{ONUKI202278}, STAR~\cite{Chen:2024aom},
CMS~\cite{Hayrapetyan_2024}, ATLAS~\cite{Aad_2024},
LHCb~\cite{doi:10.1142/S0217751X15300227}, GlueX~\cite{GlueX:2020idb},
etc.  These spectrometers are designed to measure the four-momenta of
stable final-state particles, including $e^{\pm}$, $\mu^{\pm}$,
$\pi^{\pm}$, $K^{\pm}$, $p^{\pm}$, and $\gamma$. For these
final-state particles, the spin information is normally not measured
since typically it would require an external
polarimeter~\cite{NakagawaAIP, PhysRevD.79.094014, ComptonHERA,
ComptonJLab, MottJLab2020}. These devices either have restricted
angular acceptance or cannot be accommodated in collider
spectrometers. This has made the information on the final-state
incomplete, limiting experimental studies in nuclear and particle
physics.

A novel technique has been proposed to measure final-state nucleon
polarization in large acceptance spectrometers, without sacrificing
traditional detector functions~\cite{c642-1lzb}. The essence of the
technique is to identify a proper scattering layer, which usually exists
in most spectrometers. Using the existing tracking detector and
scattering layer, most general-purpose spectrometers have the
potential to function as large-acceptance polarimeters without
hardware modification.

The measurement of spin polarization in experiments involving initial-state
polarized particles utilizes the asymmetry in the angular
distribution of scattered final-state particles. For a two-body
reaction where a spin-$\frac{1}{2}$ particle is transversely
polarized, the scattering differential cross section is generally
expressed as~\cite{PhysRevLett.90.142301, Bystricky:1976jr}
\begin{equation} \frac{{\rm d}^2\sigma}{{\rm d}\phi \, {\rm d}\!\cos\theta} =
  \frac{1}{2\pi}\frac{{\rm d}\sigma_0}{
    {\rm d}\cos\theta}\left[1+\mathcal{P}_yA_N\!(\theta)\cos\phi\right].
\end{equation}
Here, $\sigma_0$ represents the unpolarized cross
section. $\theta$ and $\phi$ denote the polar and azimuthal angles of
the emitted particles in the center of mass frame of the scattering
with the transverse polarization axis pointing to the $y$-axis.
$\mathcal{P}_y$ corresponds to the transverse polarization to be
determined. The analyzing power, $A_N(\theta)$, which depends both on energy
scattering angle, has been measured
extensively~\cite{PhysRev.148.1289,PhysRev.163.1470, PhysRev.95.1348,
ALBROW1970445, PhysRevC.24.1778, vonPrzewoski:1998ye,
PhysRevLett.41.384, PhysRevD.21.580, PhysRevD.40.35, PhysRev.105.288,
GREENIAUS1979308}.  The transverse polarization $\mathcal{P}_y$ can be
obtained from the final azimuthal distribution as long as
$A_N(\theta)$ is known.  A wealth of $A_N$ data have been measured for
the proton-proton $pp$ and proton-carbon $p\textrm{C}$ elastic
scattering processes~\cite{PhysRev.148.1289,PhysRev.163.1470,
PhysRev.95.1348, ALBROW1970445, PhysRevC.24.1778, vonPrzewoski:1998ye,
PhysRevLett.41.384, PhysRevD.21.580, PhysRevD.40.35, PhysRev.105.288,
GREENIAUS1979308}, and $pp$ and $p\textrm{C}$ scattering is
widely used for accelerator beam polarimeters.  In several fixed-target
experiments~\cite{PhysRevLett.88.092301}, the final-state polarization
of the proton was also measured with these reactions.

In this paper, we analyze the scattering asymmetry of $pp$ scattering
using polarized protons from hyperon decay at BESIII and determine the
average value of $\mathcal{P}_y A_N$. The measured value is consistent
with the expected value, which validates the function of the BESIII
detector as a precision large-acceptance proton polarimeter.

\section{BESIII experiment}

The BESIII detector records symmetric $e^+e^-$ collisions provided by
the BEPCII storage ring operating in the center-of-mass (c.m.) energy
range $\sqrt{s}=1.84-4.95$~GeV. BESIII has collected
$10.09\times10^{9}$ $J/\psi$ events~\cite{BESIII:2021cxx} including
processes such as $J/\psi\to \Lambda\bar \Lambda$ with branching
fraction $1.88(8)\times10^{-3}$~\cite{PhysRevD.110.030001}.  The
BESIII detector consists of a helium-based multilayer drift
chamber~(MDC), a plastic scintillator time-of-flight system (TOF), and
a CsI(Tl) electromagnetic calorimeter~(EMC), which are all enclosed in
a superconducting solenoidal magnet providing a 1.0~T magnetic
field. The momentum resolution of charged particle at $1~{\rm GeV}/c$
is $0.5\%$. The details are described in
Refs.~\cite{ABLIKIM2010345,CAO2020163053}.

A layer of high-purity mineral oil with a thickness of 0.8~mm
circulates as the coolant between the inner and outer beryllium shells
of the beam pipe. The inner radius of the oil layer is
32.3~mm~\cite{ABLIKIM2010345, PhysRevLett.127.012003,
Dai:2024myk}. The mineral oil is composed of carbon and hydrogen,
perfect for nucleon polarization measurements. The tracking detector
has an inner wall made of carbon fiber with a thickness of 1.2 mm and
an inner radius of 63.0~mm. It contains carbon, hydrogen and oxygen,
which are also good target candidates. The coolant and the supporting
structure were originally minimized in order not to impact the
detector performance. The details on the mineral oil and inner wall of
MDC are described in Ref.~\cite{Dai:2024myk}. While carbon is an
effective material for analyzing proton polarization, its use in
$p\textrm{C}$ scattering at BESIII presents technical challenges for
background suppression, as the low-momentum recoil carbon nuclei
cannot be detected.  Here, we only utilize the hydrogen element to
measure the polarization of final-state protons and antiprotons. With
hydrogen as the target, both the target proton and the scattered
proton (anti-proton) are detected.

\section{Polarized proton or antiproton from hyperon decay}

\begin{figure}
\includegraphics[width=0.46\textwidth]{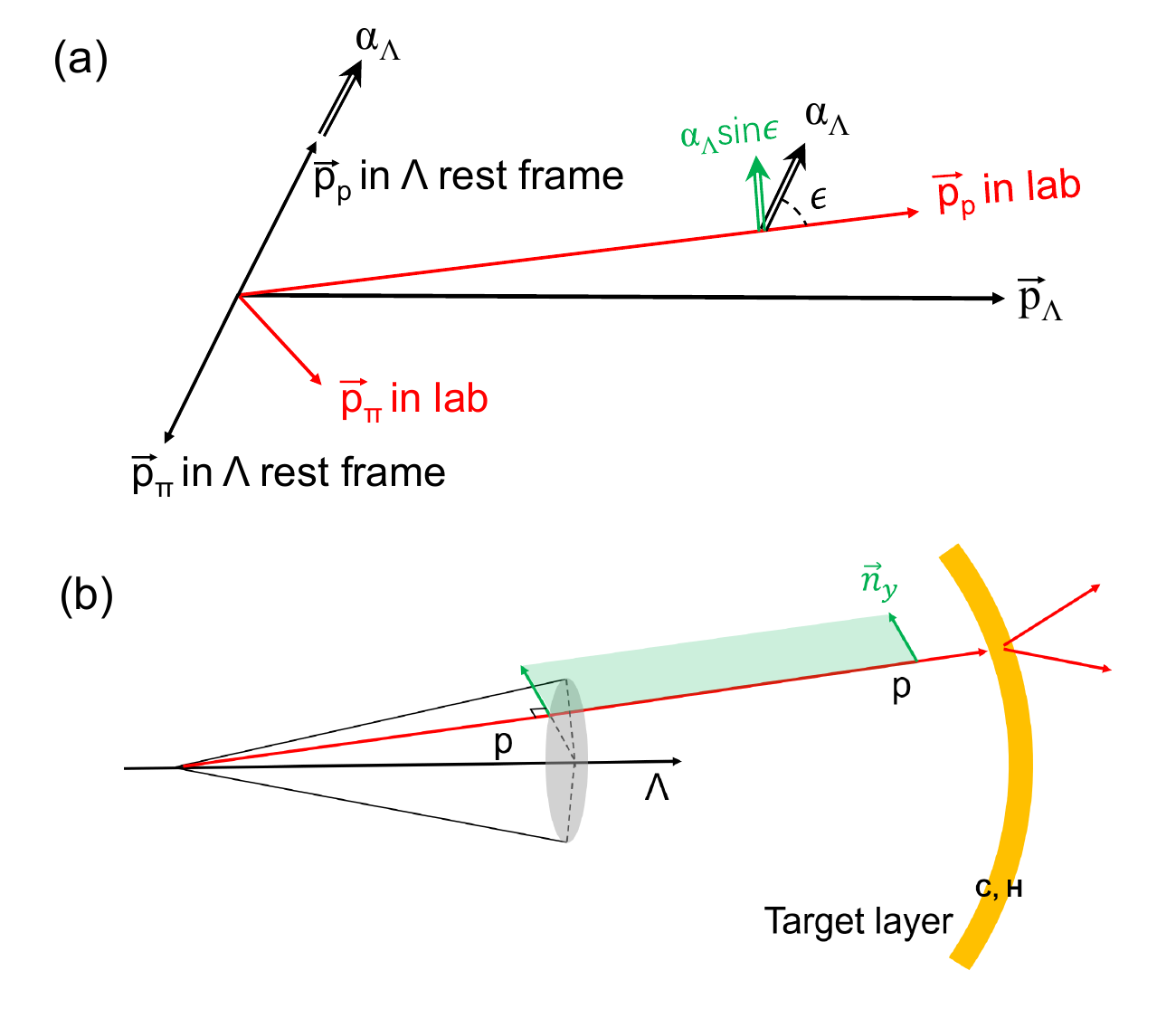}
\caption{\label{fig:p_from_lambda} Principle of this experiment. Panel
(a) shows how polarized protons are produced in $\Lambda$ weak decays:
$\alpha_{\Lambda}\sin\epsilon$ (in green color) indicates the transverse
component of proton polarization, and $\epsilon$ is the angle between the
momentum vector and the spin direction of the proton in the laboratory
frame. Panel (b) demonstrates how proton polarization is measured via
$pp$ scattering: $\vec{n}_y$ indicates the transverse polarization
axis defined in the laboratory frame. } \end{figure}

To validate our method, protons with large and known transverse
polarization are needed. For BESIII and most experiments with
sufficient energy to produce hyperons, the protons from hyperon decays
are ideal candidates. For example, protons from the $\Lambda$ decay
($\Lambda \rightarrow p\pi^-$) are longitudinally polarized in the
rest frame of $\Lambda$ with the polarization parameter
$\alpha_\Lambda=0.746\pm0.008$~\cite{PhysRevLett.129.131801,
PhysRevD.110.030001}. As depicted in Fig.~\ref{fig:p_from_lambda}(a),
these protons will naturally acquire transverse polarization when
boosted into the laboratory frame. The transverse component of proton
polarization, shown in green in Fig.~\ref{fig:p_from_lambda}(a), is
$\alpha_{\Lambda}\sin\epsilon$, where $\epsilon$ is the angle between
the momentum vector and the spin direction of the proton in the
laboratory frame.  Figure~\ref{fig:p_from_lambda}(b) shows how the
polarization of the proton from $\Lambda$ decay is measured by $pp$
elastic scattering with ``target'' protons in the spectrometer
material. Here, $\vec{n}_y$ indicates the transverse polarization
axis. Thus, event-by-event determination of the proton transverse
polarization axis is required. The coordinate system of the scattering
process is defined as follows: The $z$-axis points along the direction
of the incident proton (proton from Lambda decay). The $y$-axis is
aligned with $\vec{n}_y$, which is perpendicular to the $z$-axis and
lies in the decay plane formed by the proton and Lambda. The $x$-axis
is then chosen such that the coordinate forms a right-handed
system. The polar and azimuthal angles of the scattered proton are
defined within this scattering coordinate system.

The transverse polarization of the proton is affected by the hyperon
momentum only negligibly; it mainly depends on the angle of proton
emission in the hyperon rest frame~\cite{c642-1lzb}. This provides a
reliable way to determine the polarization by means of the proton
emission angle. In fact, polarized proton and antiproton beams at
Fermilab are prepared in this manner~\cite{GROSNICK1990269}. At
BESIII, protons in the acceptance of the spectrometer from $\Lambda$
decay with a $\Lambda$ momentum around 1 GeV/$c$ have an average
transverse polarization of $0.59 \pm 0.01$~\cite{c642-1lzb}. For the decay
of $\Sigma^+$ to $p\pi^0$, the final-state proton acquires a larger
transverse polarization of about $0.75 \pm 0.01$, since the initial
(longitudinal) polarization (in the rest frame of $\Sigma^+$) is
greater
($|\alpha_{\Sigma^{\pm}}|=0.99\pm0.01$)~\cite{PhysRevD.110.030001}.

\section{Data analysis}

\begin{figure*}
\includegraphics[width=0.8\textwidth]{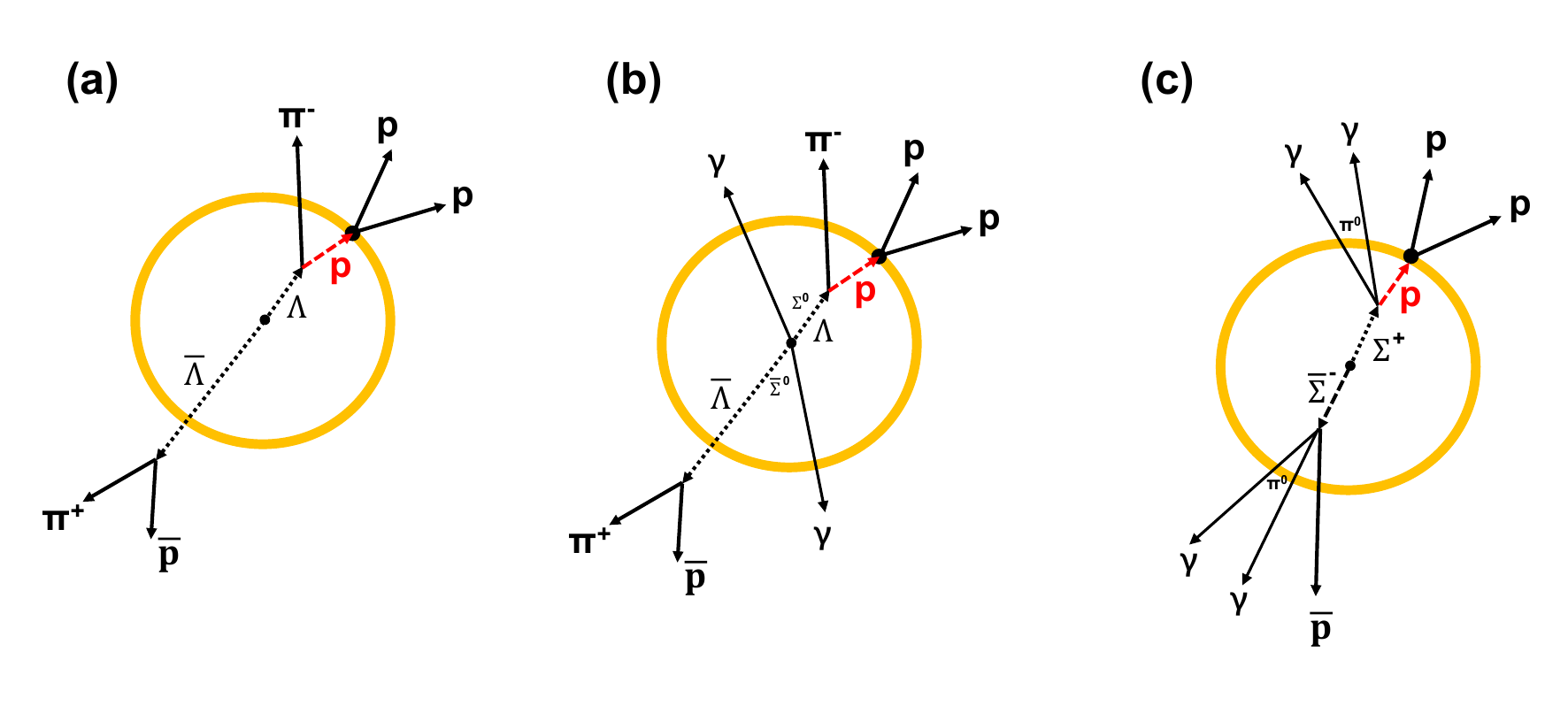}
\caption{\label{fig:topo_4_channel} {Topology of signal events (a)
$e^{+}e^{-}\rightarrow J/\psi \rightarrow \Lambda\bar{\Lambda}$, (b)
$e^{+}e^{-}\rightarrow J/\psi \rightarrow \Sigma^0\bar{\Sigma}^0$ and
(c) $e^{+}e^{-}\rightarrow J/\psi \rightarrow \Sigma^+\bar{\Sigma}^-$
 followed by elastic $pp$ scattering. The yellow circle denotes the
``target'' layer. The red arrows represent the polarized protons from
hyperon decay.  } } \end{figure*}

Three decays of the $J/\psi$ meson into hyperon pairs are analyzed:
$J/\psi\rightarrow \Lambda\bar{\Lambda}$, $\Sigma^0\bar{\Sigma}^0$,
and $\Sigma^+\bar{\Sigma}^-$. The reaction of $J/\psi \rightarrow
\Lambda\bar{\Lambda}$ is identified by the subsequent decays
$\Lambda\rightarrow p\pi^{-}$ and
$\bar{\Lambda}\rightarrow\bar{p}\pi^{+}$. This process yields four
charged tracks: $p$, $\pi^-$, $\bar{p}$ and $\pi^+$, and a candidate
event satisfying this condition is shown in
Fig.~\ref{fig:topo_4_channel}(a). The proton scatters elastically, and
an additional charged track, identified as a proton, is also present,
consistent with being knocked out of the target material by the proton
from the $\Lambda$ decay.  In the $J/\psi \rightarrow
\Sigma^0\bar{\Sigma}^0$ channel, we analyze the decay chain
$\Sigma^0\rightarrow \gamma\Lambda \rightarrow \gamma p\pi^-$ and
$\bar{\Sigma}^0\rightarrow \gamma\bar{\Lambda} \rightarrow \gamma
\bar{p}\pi^+$, where protons (antiprotons) inherit polarization from
the $\Lambda$ ($\bar{\Lambda}$) decay. Such events contain five
charged particles and two photons
(Fig.~\ref{fig:topo_4_channel}(b)). The $J/\psi \rightarrow
\Sigma^+\bar{\Sigma}^-$ channel features $\Sigma^+$ ($\bar{\Sigma}^-$)
decaying to $p$ ($\bar{p}$) and $\pi^0$, yielding three charged
particles and four photons in the final state
(Fig.~\ref{fig:topo_4_channel}(c)). Although not shown in
Fig.~\ref{fig:topo_4_channel}, the antiproton-proton $\bar{p}p$
elastic scattering is also studied.

Charged tracks are reconstructed in the MDC within the angular region
$\vert\!\cos\theta_\text{lab}\vert< 0.93$, where $\theta_\text{lab}$ is
the polar angle of the track with respect
to the $z$ axis in the laboratory frame which is the symmetry axis of
the detector. Particle identification~(PID) for charged tracks
combines measurements of the energy deposited in the MDC~(d$E$/d$x$)
and the flight time in the TOF to form likelihoods
$\mathcal{L}(h)~(h=p,K,\pi)$ for each hadron $h$ hypothesis. Tracks
are identified as protons when the proton hypothesis has the greatest
likelihood, ($\mathcal{L}(p)>\mathcal{L}(K)$ and
$\mathcal{L}(p)>\mathcal{L}(\pi)$), while charged pions are identified
when $\mathcal{L}(\pi)>\mathcal{L}(K)$ and
$\mathcal{L}(\pi)>\mathcal{L}(p)$. Neutral candidates are identified
using electromagnetic showers in the EMC. The deposited energy of each
shower must be larger than 25 MeV in the barrel region ($\vert\!\cos\theta_\text{lab}\vert<
0.8$) and larger than 50 MeV in the end-cap region ($0.86 <
\vert\!\cos\theta_\text{lab}\vert< 0.92$). To suppress electronic noise and showers
unrelated to the event, the difference between the EMC time of the
neutral candidate and the event start time is required to be within
[0, 700] ns.

\begin{figure}
\includegraphics[width=0.45\textwidth]{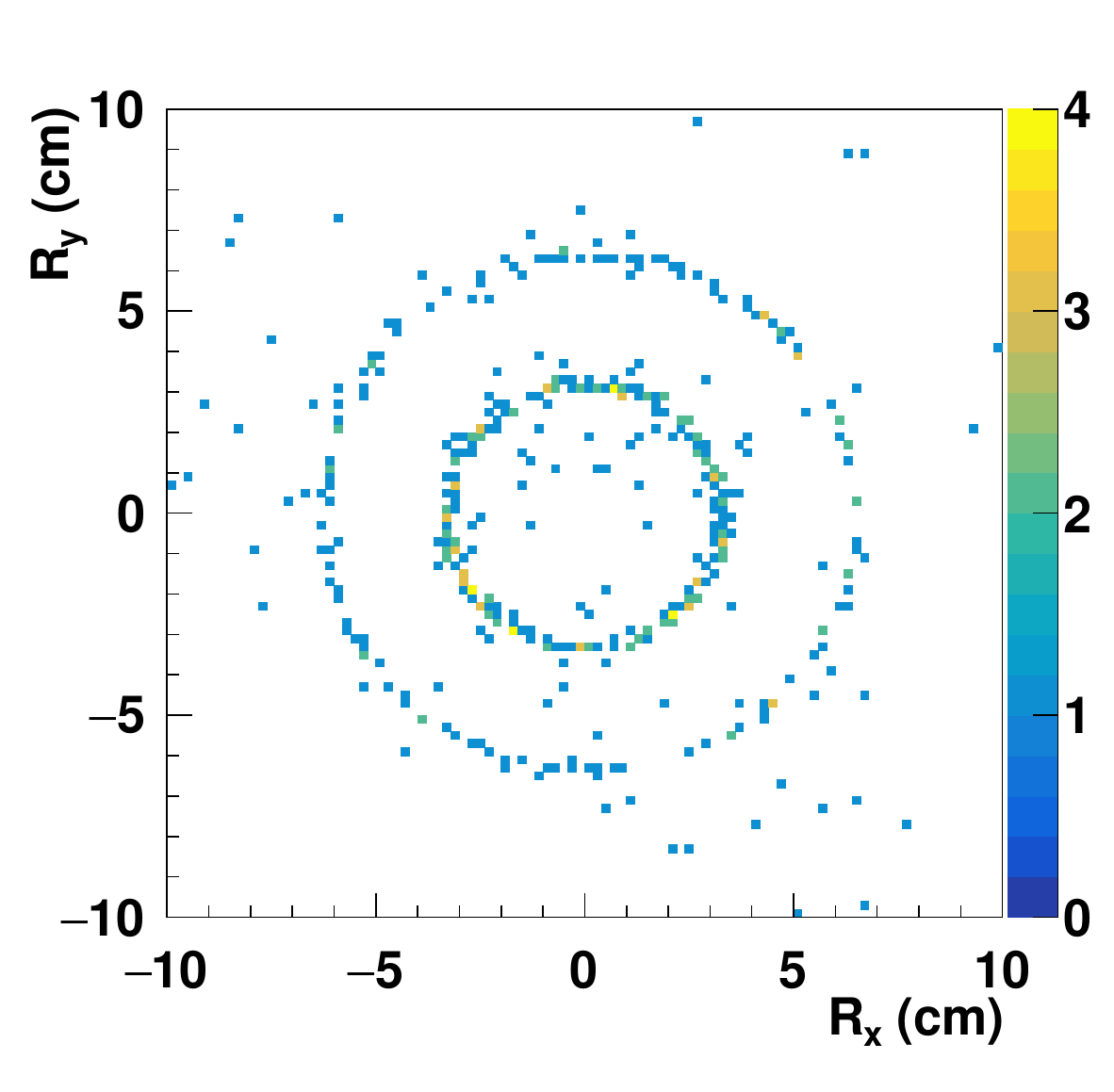}
\caption{\label{fig:Rxy_pp} {Distribution of $pp$ scattering vertices
at the beam pipe (smaller ring) and inner wall of the MDC (larger ring)
for the process $J/\psi \rightarrow \Lambda \bar{\Lambda} \rightarrow
p\bar{p}\pi^+\pi^-$ of experimental data.}  } \end{figure}

The polarization measurement makes use of the elastic scattering
between protons (antiprotons) and target protons in the cooling and
supporting material.  BESIII has demonstrated an excellent
reconstruction capability for such scattering events, as evidenced by
previous hyperon--nucleon scattering studies that successfully employed
secondary interactions in the beam
pipe~\cite{BESIII:2023clq,BESIII:2024geh}. Figure~\ref{fig:Rxy_pp}
displays the vertex distribution for $pp_{\rm{pipe}}$ scattering
events from the process $J/\psi \rightarrow \Lambda \bar{\Lambda}
\rightarrow p\bar{p}\pi^+\pi^-$ followed by $p
p_{\rm{pipe}}\rightarrow p p_{\rm{pipe}}$. Here, $p_{\rm{pipe}}$
represents the proton from the beam pipe or the inner wall of the MDC. The
image shows that scatterig from the beam pipe and the MDC inner
wall clearly are resolved.  Since scattering occurs before tracks reach the MDC, we
determine the incident proton (antiproton) momentum through recoil
kinematics. For illustration, considering the topology shown in 
Fig.~\ref{fig:topo_4_channel}(a), the unmeasured momentum
of the proton to be scattered is reconstructed via $p_p = p_{e^+} +
p_{e^-} - p_{\pi^+} - p_{\bar{p}} - p_{\pi^-}$, where, $p_{e^{\pm}}$
represent the beam four momenta. In this case, the momentum resolution of the recoil proton is approximately 8 MeV/$c$, and the angular resolutions for $\theta$ and $\phi$ resolutions are about 7 mrad, both of which are sufficiently precise. We also calculate the target proton
momentum ($p_{\rm{target}}$) using four-momentum conservation between
final-state and incident protons, and require the momentum
$|p_{\rm{target}}| < 50$~MeV/$c$.  This requirement suppresses the
background collision process where the incident proton hits a proton in a
carbon or oxygen nucleus. In this background process, the target proton
is not at rest as a free hydrogen target, but has a sizable Fermi
momentum. This collision is referred to as a ``quasi-free nuclear
collision" in this context.

\begin{figure*}[htb]
\centering
\includegraphics[width=0.95\textwidth]{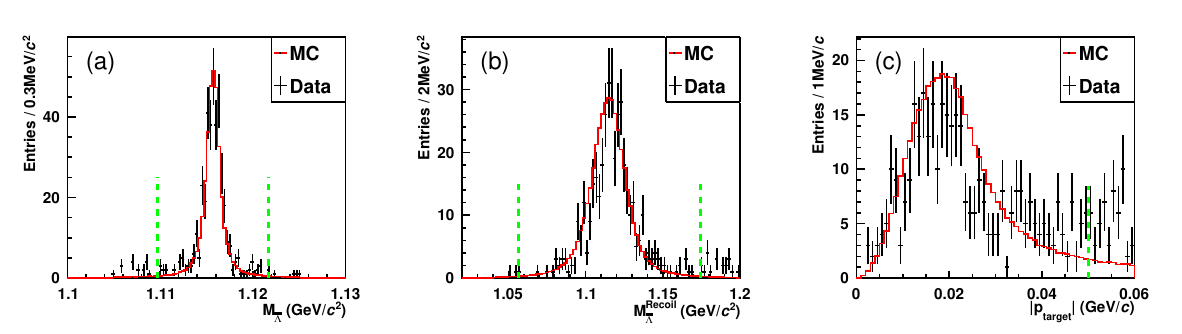}
\caption{Event selection of $pp$ elastic scattering from the process
$e^{+}e^{-}\rightarrow J/\psi \rightarrow \Lambda\bar{\Lambda}$. Plot
(A) shows the invariant mass of the tagged $\bar{\Lambda}$. Plot (B)
is the recoil mass of the tagged $\bar{\Lambda}$. Plot (C) shows the
calculated momentum of the target proton. The red histograms are
signal MC, the black points with error bars are the data, and the
vertical lines indicate the applied requirements.  } 
\label{fig:Sig_pp_Lambda} \end{figure*}

For the process $e^{+}e^{-}\rightarrow J/\psi \rightarrow
\Lambda\bar{\Lambda}$, Fig.~\ref{fig:Sig_pp_Lambda} presents the
selection of $pp$ scattering events, with Monte Carlo (MC) simulations
shown in red and experimental data represented by black points with
error bars. The signal MC event generation is carried out in two
stages. First, the decay processes $J/\psi\rightarrow
\Lambda\bar{\Lambda} \rightarrow p\pi^- \bar{p}\pi^+$, are generated
using the standard {\sc EvtGen} generator~\cite{LANGE2001152,
PingRong-Gang_2008}. Subsequently, the produced proton (anti-proton)
is transported to the beam pipe or the inner wall of MDC,
where it undergoes elastic scattering with a target proton. This transport
includes both deflection of the track and the spin precession in the
magnetic field of the BESIII detector.  Finally, the
resulting final-state particles are processed by a {\sc Geant4}
based software package~\cite{AGOSTINELLI2003250}.  In Fig.~\ref{fig:Sig_pp_Lambda}, vertical green
lines indicate the mass requirement, defined as a
$\pm 4\sigma$ window, determined using MC
samples. Figure~\ref{fig:Sig_pp_Lambda}(a) shows the invariant mass of
the tagged $\bar{\Lambda}$, and Fig.~\ref{fig:Sig_pp_Lambda}(b) shows
the recoil mass of the tagged $\bar{\Lambda}$, where a clear peak at
the $\Lambda$ mass is seen. Figure~\ref{fig:Sig_pp_Lambda}(c) shows
the calculated momentum of the recoiled target proton, where a
difference between data and signal MC is seen in the higher mass
region due to the background contribution from quasi-free nuclear
collisions. To suppress this background, the momentum is required to
be less than 50 MeV/$c$.

To suppress backgrounds, we require that the $\Lambda$ decays ahead of
the scattering of the final-state proton.  This is achieved by the
requirement $R_{xy}^{pp}-R_{xy}^{pp\pi} > 0$, where $R_{xy}^{pp}$ and
$R_{xy}^{pp\pi}$ are the distances in the transverse plane between the
collision point and the reconstructed vertex of  $pp_{\rm{pipe}}$ and
the $\Lambda$ decay vertex, respectively.

\begin{figure*}[htb]
\centering
\includegraphics[width=0.95\textwidth]{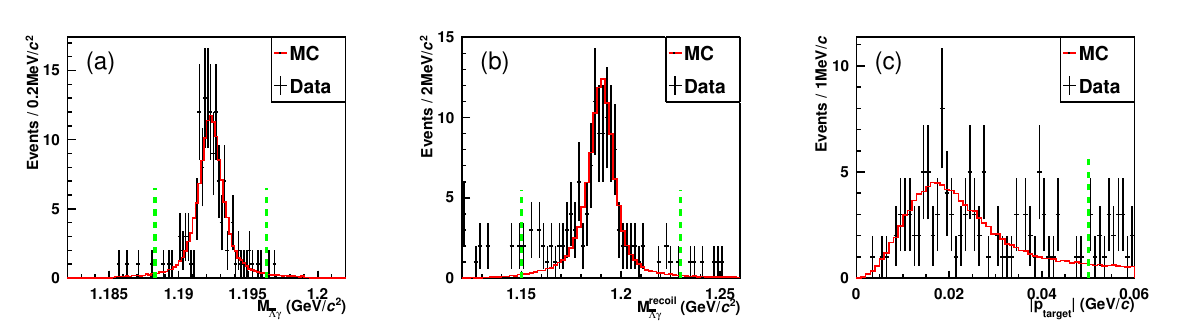}
\caption{Event selection of $pp$ elastic scattering from the process
$e^{+}e^{-}\rightarrow J/\psi \rightarrow
\Sigma^0\bar{\Sigma}^0$. Plot (A) shows the invariant mass of the
tagged $\bar{\Sigma}^0$. Plot (B) is the recoil mass of the tagged
$\bar{\Sigma}^0$. Plot (C) shows the calculated momentum of the target
proton. The red histograms are signal MC, the black points with error
bars are the data, and the vertical lines indicate the applied
requirements.  } \label{fig:Sig_pp_Sigma0} \end{figure*}

\begin{figure*}[htb]
\centering
\includegraphics[width=0.95\textwidth]{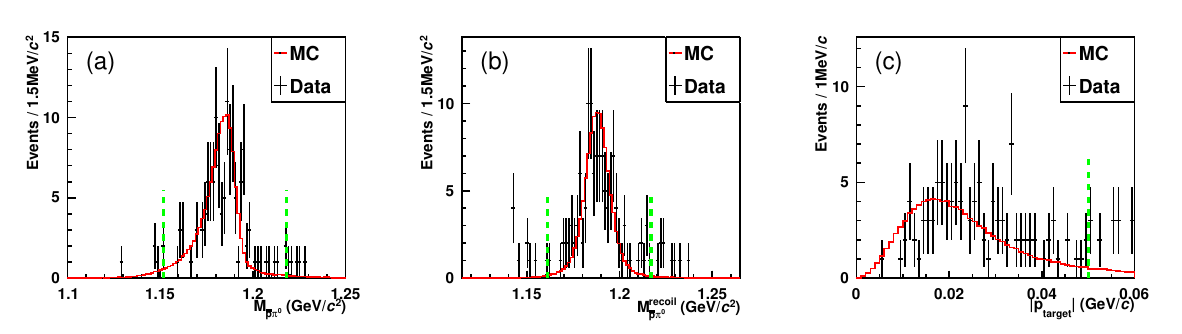}
\caption{Event selection of $pp$ elastic scattering from the process
$e^{+}e^{-}\rightarrow J/\psi \rightarrow
\Sigma^+\bar{\Sigma}^-$. Plot (A) shows the invariant mass of the
tagged $\bar{\Sigma}^-$. Plot (B) is the recoil mass of the tagged
$\bar{\Sigma}^-$. Plot (C) shows the calculated momentum of the target
proton. The red histogram is signal MC, the black points with error
bars are experimental data. Vertical lines indicate the applied
requirements.  } \label{fig:Sig_pp_Sigma} \end{figure*}

For the $pp_{\rm{pipe}}$ scattering from $e^{+}e^{-}\rightarrow
J/\psi\rightarrow\Sigma^0\bar{\Sigma}^0$, the invariant mass of
$\bar{\Sigma}^0$, constructed with a $\bar{\Lambda}$ and a photon, is
shown in Fig.~\ref{fig:Sig_pp_Sigma0}(a). The recoil mass of the
tagged $\bar{\Sigma}^0$ is shown in Fig.~\ref{fig:Sig_pp_Sigma0}(b),
where a clear peak at the $\Sigma^0$ mass is
seen. Figure~\ref{fig:Sig_pp_Sigma0}(c) shows the calculated momentum
of the recoiling target proton. For the $pp_{\rm{pipe}}$ scattering
from $e^{+}e^{-}\rightarrow J/\psi\rightarrow\Sigma^+\bar{\Sigma}^-$,
a tagged $\bar{\Sigma}^-$ is constructed with a $\bar{p}$ and a
$\pi^0$ which is reconstructed by a pair of photons selected by a
kinematic fit with a constraint on the $\pi^0$ mass. The mass of the
tagged $\bar{\Sigma}^-$ is shown in
Fig.~\ref{fig:Sig_pp_Sigma}(a), and the recoil mass of the tagged
$\bar{\Sigma}^-$ is shown in Fig.~\ref{fig:Sig_pp_Sigma}(b), where a
clear peak at the $\Sigma^+$ mass is
seen. Figure~\ref{fig:Sig_pp_Sigma}(c) shows the calculated momentum
of the recoiling target proton.

\section{\boldmath Expected and measured $\mathcal{P}_yA_N$}

\begin{figure}[htbp]
\centering
\includegraphics[width=0.45\textwidth]{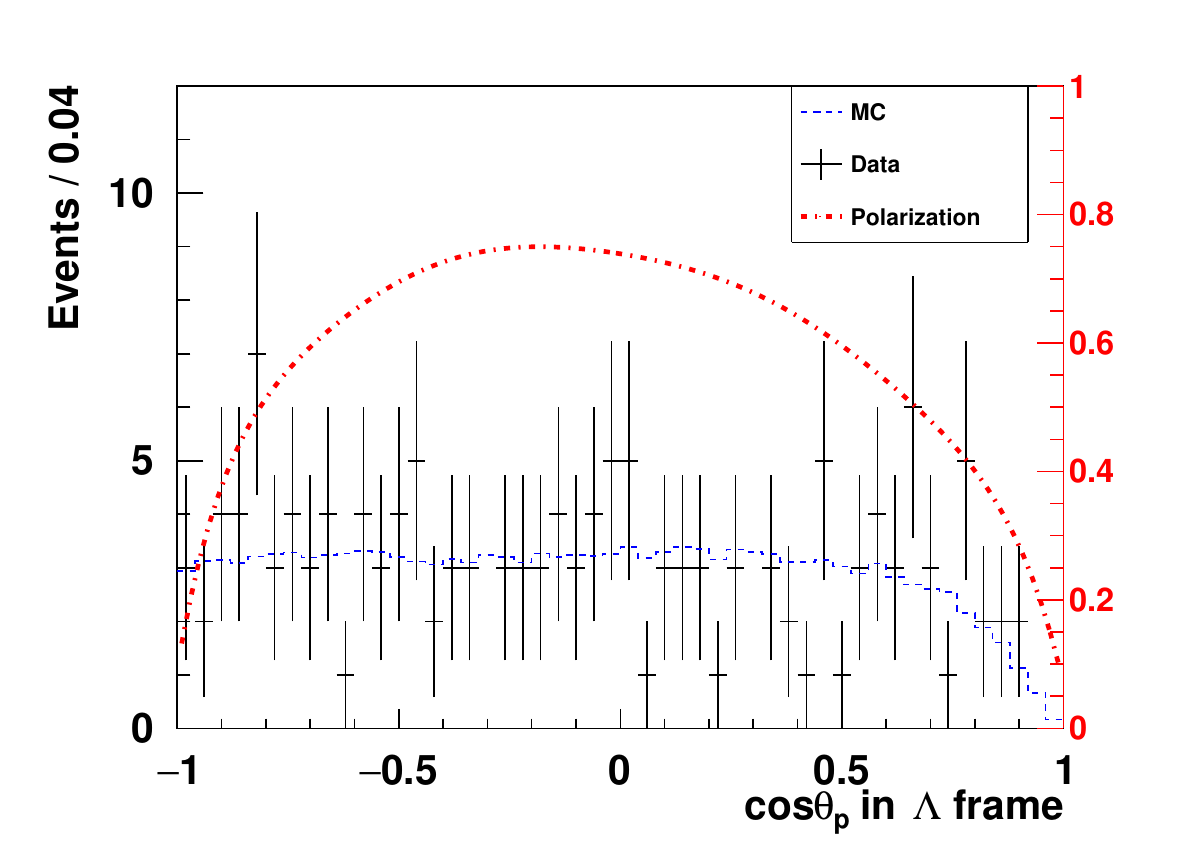}
\caption{Polar angle distribution of the proton in the $\Lambda$ rest frame
 from the process $e^{+}e^{-}\rightarrow J/\psi \rightarrow
\Lambda\bar{\Lambda}$. The blue dashed histogram is signal MC, the
black points with error bars are data. The red dot-dashed line is
the transverse polarization curve with axis at the right side.}
\label{fig:pol_t_expected} \end{figure}

\begin{figure}[htbp] \centering
\includegraphics[width=0.45\textwidth]{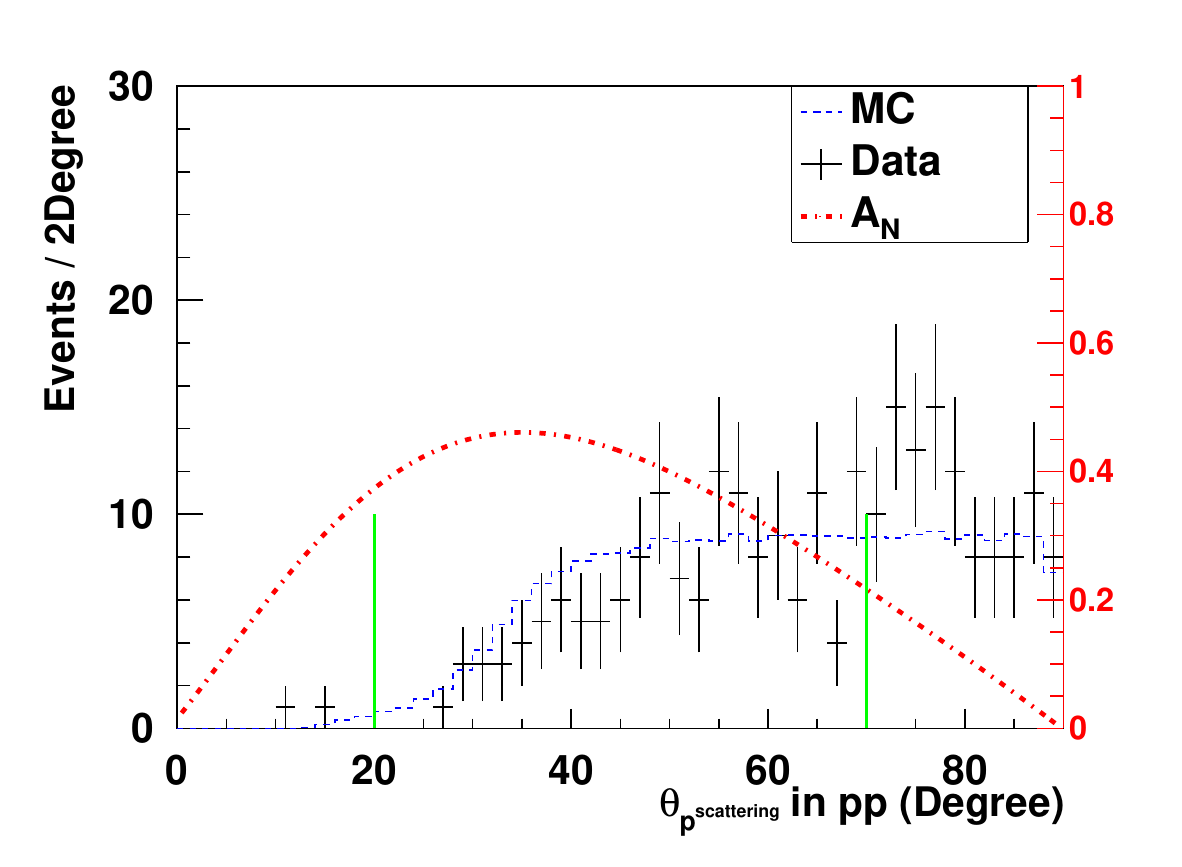}
\caption{Polar angle distribution of the scattered proton in the $pp$
c.m. frame from the process $e^{+}e^{-}\rightarrow J/\psi \rightarrow
\Lambda\bar{\Lambda}$. The black points with error bars are data. The
blue dashed histogram is signal MC. The red dot-dashed line is the
analyzing power curve with axis at the right side. The two vertical
green lines indicate the kinematic region with high analyzing power.}
\label{fig:An_expected} \end{figure}

\begin{table*}[t]
\renewcommand{\arraystretch}{1.6}
\setlength{\tabcolsep}{18pt} 
\centering
\caption{The expected average $\mathcal{P}_yA_N$ of $p$ ($\bar{p}$)
from MC simulation by channel, the weighted average, and the number of
events satisfying selection requirements in data. The numbers in
bracket are for the anti-proton. }

\begin{tabular}{ c | c | c } \hline
Channel & $\mathcal{P}_yA_N$ & $N_{\rm sig}$ \\\hline
$J/\psi\rightarrow \Lambda\bar{\Lambda}$ &  $0.23\pm0.01$ ($0.20\pm0.01$) & 143 (176) \\
$J/\psi\rightarrow\Sigma^0\bar{\Sigma}^0$ & $0.20\pm0.01$ ($0.20\pm0.01$) & 42 (71) \\
$J/\psi\rightarrow\Sigma^+\bar{\Sigma}^-$ & $0.26\pm0.01$ ($0.24\pm0.01$) & 43 (59) \\\hline
Average & $0.23\pm0.01$ ($0.21\pm0.01$) & \\ \hline
\end{tabular} \label{table:pol_t_summary}
\end{table*}

The distribution of the proton emission angle in the
$\Lambda$ rest frame, $\theta_\textrm{p}^{\Lambda}$, from the process
$e^{+}e^{-}\rightarrow J/\psi \rightarrow \Lambda\bar{\Lambda}$ is plotted in
Fig.~\ref{fig:pol_t_expected}. The black points with error bars are
the experimental data, the blue dashed histogram is the MC simulation. The red
dot-dashed line is the proton transverse polarization as a function of
the emission angle, with its value shown by the axis on the right.

To determine the polarization, the analyzing power is required.  The
$\vec{p}p$ analyzing power is retrieved from the SAID (Scattering
Analyses Interactive Dial-in) database~\cite{PhysRevC.62.034005,
*PhysRevC.76.025209, *GWDAC}, where all $pp$ spin observables are
collected and analyzed. Although polarized antiproton experiments are
scarce, measurement at LEAR~\cite{KUNNE1988557} provides the
$\vec{\bar{p}}p$ analyzing power in our kinematical region.  In
Fig.~\ref{fig:An_expected}, the $pp$ analyzing power at 0.9 GeV/$c$
from the SAID database is presented by the red dot-dashed curve as a
function of the scattering angle, with its value shown by the axis on
the right. The BESIII experimental distribution and MC simulation from
the process $e^{+}e^{-}\rightarrow J/\psi \rightarrow
\Lambda\bar{\Lambda}$ are shown with black dots with error bar and
blue dashed curve, respectively. Events beyond the central angular range
are discarded as indicated by the vertical lines, to achieve a large
asymmetry.

Listed in Table~\ref{table:pol_t_summary} are the average
$\mathcal{P}_yA_N$ values, determined from MC simulation, for $pp$ and
$\bar{p}p$ scattering for the three hyperon decay channels, the
numbers of signal events passing selection criteria in data, and the
average $\mathcal{P}_yA_N$ weighted by the numbers of signal events.
Combining both the $pp$ and $\bar{p}p$ scattering samples, the
expected value of the average polarization times the analyzing power
is $\left<\mathcal{P}_yA_N\right>_{\rm exp} = 0.22 \pm 0.01$.

\begin{figure}
\includegraphics[width=0.45\textwidth]{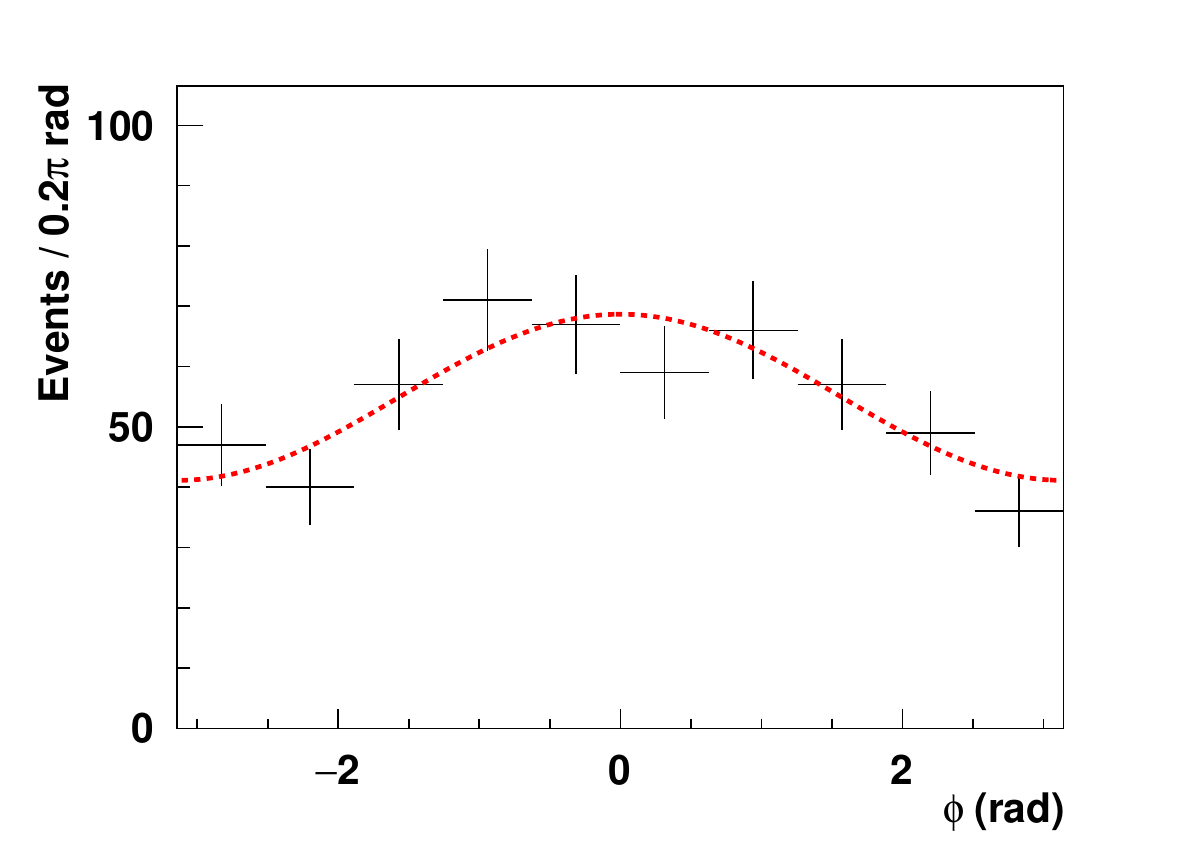}

\caption{\label{fig:SSA_validation} {Azimuthal distribution of $pp$
and $\bar{p}p$ elastic scattering for experimental data. Both $pp$ and
$\bar{p}p$ events, induced by polarized protons and antiprotons from
$\Lambda(\bar{\Lambda})$, $\Sigma^0(\bar{\Sigma}^0)$ and
$\Sigma^+(\bar{\Sigma}^-)$ hyperon decay processes, are combined. An
unbinned maximum likelihood fit (dashed line) yields
$\left<\mathcal{P}_yA_N\right>_{\rm scat} = 0.243 \pm 0.059$. }}
\end{figure}

The experimental azimuthal distribution of scattered protons and
antiprotons is shown in Fig.~\ref{fig:SSA_validation}. An unbinned
maximum likelihood fit to the azimuthal distribution is performed with
the angular distribution $\mathcal{W}(\phi; \mathcal{P}_yA_N) \propto
(1+\mathcal{P}_yA_N\cos\phi)$. The likelihood function is defined as:
\begin{equation} \begin{aligned} &\mathcal{L} =
\prod_{i=1}^{N}\mathcal{P}(\phi^{i};\mathcal{P}_yA_N)
 = \prod_{i=1}^{N}\mathcal{CW}(\phi^{i};
 \mathcal{P}_yA_N)\epsilon(\phi^{i}).
 \label{eq:maximum_fit_function_3} \end{aligned} \end{equation} Here,
 $\mathcal{P}(\phi^{i};\mathcal{P}_yA_N)$ is the probability density
 function, and the detection efficiency is denoted by
 $\epsilon(\phi^{i})$. The normalization factor $\mathcal{C}^{-1} =
 \frac{1}{N_{\rm
 MC}}\prod_{j=1}^{N_{\rm MC}}\mathcal{W}(\phi^{i};\mathcal{P}_yA_N)\epsilon(\phi^{i})$
 is estimated with the $N_{\rm MC}$ events generated with unpolarized
 scattering ($\mathcal{P}_yA_N=0$). These MC events with zero net
 polarization are subject to the same selection criteria as the
 experimental data. To ensure sufficient statistical precision, we
 generate approximately 500 times more MC events than the accepted
 data sample. For the minimization procedure, we employ the MINUIT
 optimization package~\cite{iminuit,James:1975dr} from the CERN
 library to minimize the objective function $-\ln\mathcal{L}$. The fit
 determines a value of $\left<\mathcal{P}_yA_N\right>_{\rm scat} =0.243 \pm
 0.059$, which
 is in good agreement with the expectation of
 $\left<\mathcal{P}_yA_N\right>_{\rm exp} = 0.22 \pm 0.01$.
 This successfully validates the capability of BESIII as a
 final-state proton polarimeter. Here, the uncertainty is statistical.

\begin{figure}[htbp] \centering
\includegraphics[width=0.45\textwidth]{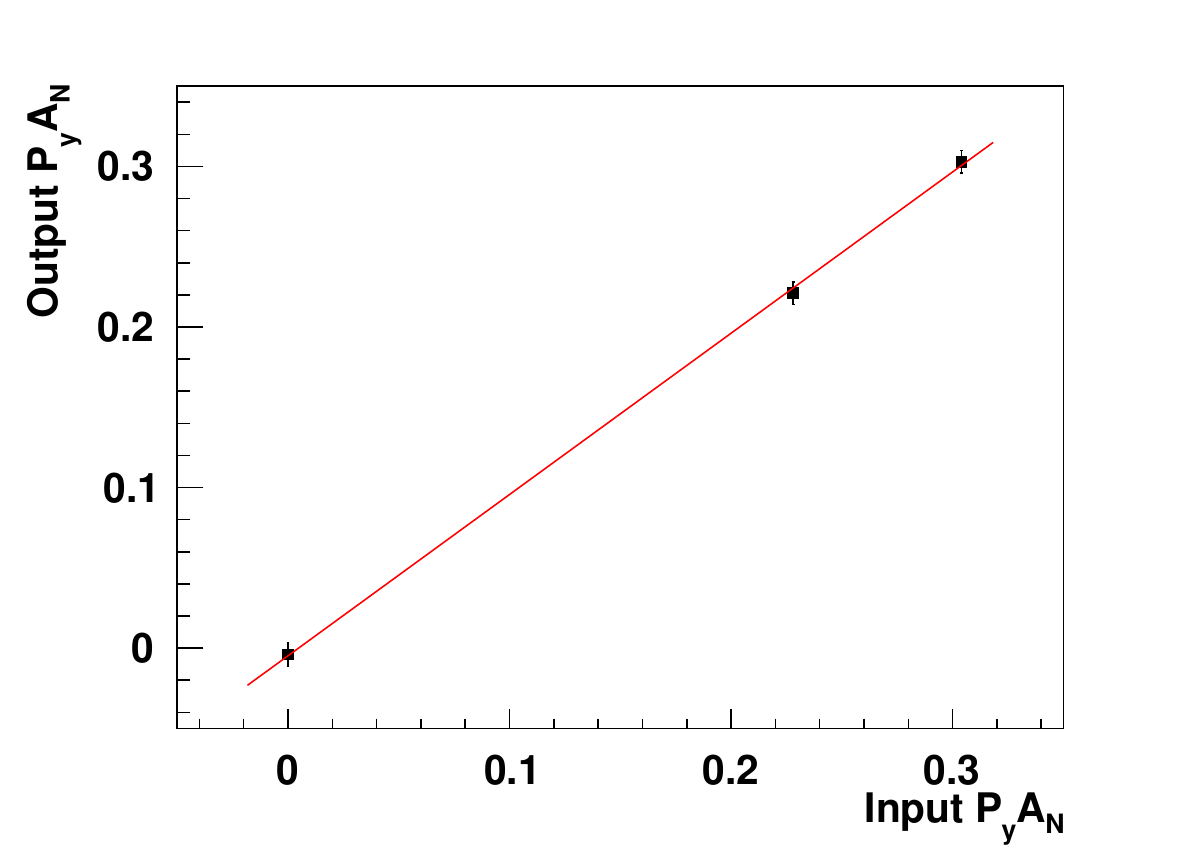}
\caption{MC checks of $pp$ scattering from $\Lambda$ decay, with
different input ${\mathcal{P}_yA_N}$ values. The output
${\mathcal{P}_yA_N}$ results are consistent with the input values. A
linear fit is performed to the results, as indicated by the red
line. } \label{fig:IO_check} \end{figure}

\begin{figure}[htbp] \centering
\includegraphics[width=0.45\textwidth]{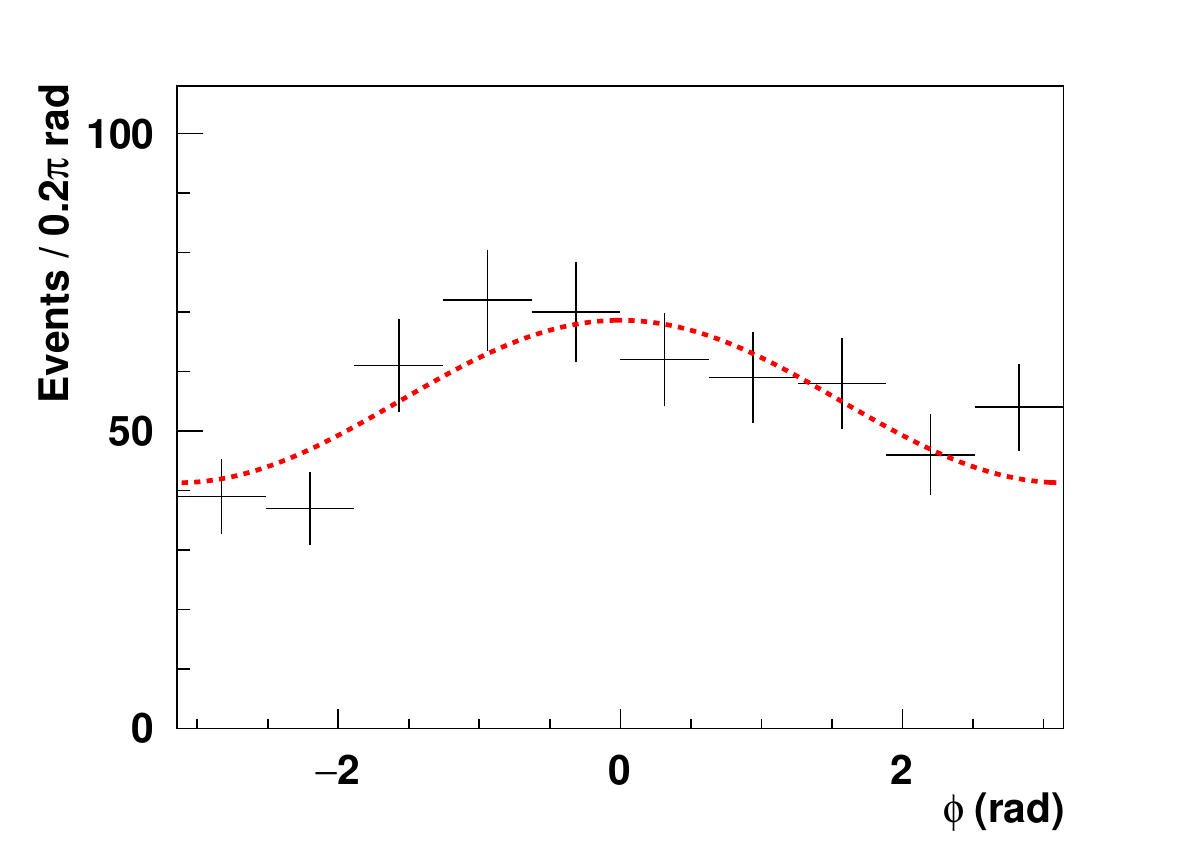}
\caption{Azimuthal distribution of $pp$ and $\bar{p}p$ elastic
scattering for a toy MC sample with same statistics as the data. The
input ${\mathcal{P}_yA_N}$ is 0.228. The black points with error bars
are the toy MC. The red dashed line is the fit result with the
obtained ${\mathcal{P}_yA_N}$ to be 0.248 $\pm$ 0.060.}
\label{fig:SSA_IO_Lambda_57percent} \end{figure}

The precision of the $\mathcal{P}_yA_N$ measurement is investigated with toy MC
samples of different input asymmetries. For the toy MC, the standard
{\sc EvtGen} generator~\cite{LANGE2001152, PingRong-Gang_2008}, is
used to generate the event prior to the scattering
process. Subsequently, the produced proton (anti-proton) is
transported to the beam pipe or the inner wall of MDC, taking into
account both the deflection of the track and the spin precession within the BESIII magnetic field. The proton scattering is then handled with different input of scattering asymmetries. Finally, the
resulting final-state particles are processed by a {\sc Geant4}
based software package~\cite{AGOSTINELLI2003250}. After the full {\sc Geant4} simulation
with the BESIII detector and applying the same analysis strategy as to
experimental data, the value for $\mathcal{P}_yA_N$ is obtained and compared to the
input values. Figure~\ref{fig:IO_check} shows the values obtained
 versus the input ones. A linear fit is performed,
with an intercept of $-0.005\pm0.007$ and a slope of $1.005\pm0.031$,
indicating good agreement.
This proves that the BESIII detector can accurately measure
proton polarization. Figure~\ref{fig:SSA_IO_Lambda_57percent} shows
the azimuthal angle distribution of one toy MC sample with same
statistics as the data. The $\left<\mathcal{P}_yA_N\right>_{\rm scat}$ of this MC
sample is $0.248 \pm 0.060$, which is consistent with the input value
of 0.228.

\section{Systematic uncertainties}

The systematic uncertainties on $\left<\mathcal{P}_yA_N\right>_{\rm scat}$ in this
analysis can be categorized into two components: 1) uncertainties from
event selection, including background estimation, tracking and PID
efficiencies; 2) fit procedure uncertainties. In the event selection,
mass windows of $\pm 4\sigma$ are applied to $\Lambda$,
$\bar{\Lambda}$, $\Sigma^{0}$, $\bar{\Sigma}^0$, $\Sigma^{+}$,
$\bar{\Sigma}^-$ and the recoil masses, and a requirement on the
scattering vertex is applied. The mass windows and the requirement of
the scattering vertex are changed between $\pm 3\sigma$ and $\pm
5\sigma$. The most significant change is taken as the uncertainty. In
the analysis, a requirement on the vertex difference
$R_{xy}^{pp}-R_{xy}^{pp\pi} > 0$ is used to reduce background. The
requirement is changed between $\pm0.5$~cm, and the most significant
change is taken as the uncertainty. The systematic uncertainty from
the kinematic fit is estimated by performing two kinematic fits, with
and without helix parameter corrections~\cite{PhysRevD.87.012002}. The
change from the kinematic fit is negligible. The requirement of
$|p_{\rm{target}}| < 50$~MeV/$c$ for target protons is varied by 10
MeV/$c$, with the observed difference in fit result taken as the
systematic uncertainty. The beam position in the MC simulation is
shifted by $\pm1\sigma$, with the resulting fit variation treated as
the systematic uncertainty. The polar angle distribution of
$pp$/$\bar{p}p$ scattering is modeled using a data-driven approach, in
which the polar angle distributions of $pp$/$\bar{p}p$ scattering in
experimental data are obtained and modeled with polynomial
functions. We generate alternate MC samples with different polynomial
orders for the angular distribution and take the largest fit variation
as the systematic uncertainty.
A potential background of 2\% is estimated based on the MC study,
which would change $\left<\mathcal{P}_yA_N\right>_{\rm scat}$ by 2\%, which is
assigned as the systematic uncertainty.

The uncertainty from tracking and PID efficiencies are studied by a
correction factor of data/MC ratios. Using the $J/\psi\rightarrow
p\bar{p}\pi^+\pi^-$ control sample, we get the data/MC ratio of PID
efficiency and tracking efficiency. The fit with the corrected MC
sample is performed 100 times with correction factors varied within
their $\pm1\sigma$ uncertainties. The deviation between the mean
corrected and nominal fit results on $\left<\mathcal{P}_yA_N\right>_{\rm scat}$
is negligible. The fit procedure uncertainty is evaluated using toy MC
samples of known polarization. The systematic uncertainty is taken as
the deviation between the input and output values of the fit. The
combined systematic uncertainty for $\left<\mathcal{P}_yA_N\right>_{\rm scat}$ is
determined to be $\pm$0.039. The contributions from each source are
listed in Table~\ref{table:sys_err_validation}.

\begin{table*}[t]
\renewcommand{\arraystretch}{1.3}
\setlength{\tabcolsep}{15pt}
\centering
\caption{Summary of systematic uncertainties.}
\begin{tabular}{ l | c | c}
\hline
 Source & Constraint variation/ & Uncertainty\\
 & Estimation method &  \\\hline
 Mass windows and vertex requirement & [$\pm 3\sigma$, $\pm 5\sigma$] & 0.020\\
 $R_{xy}^{pp}-R_{xy}^{pp\pi}$ requirement & $\pm0.5$ cm &0.022 \\
 Recoil proton momentum requirement & [40~MeV/$c$, 60~MeV/$c$] &0.021 \\
 Background impact  &  2\%  &0.005 \\ 
 Beam position  & $\pm1\sigma$ & 0.004 \\
 Polar angle distribution   & data-driven &0.005 \\
 Fit method    & MC simulation & 0.010 \\
  \hline
 Sum          & & 0.039 \\\hline  
 
\end{tabular}
\label{table:sys_err_validation}
\end{table*}

\section{Summary}

In this paper, three different hyperon decays to proton or
anti-proton are analyzed with the BESIII detector, where the hyperons
are produced in $10.09\times10^{9}$ $J/\psi$
events.  The protons and anti-protons are polarized, and some of them
scatter on hydrogen in the material of the detector, close to the
production point of the (anti-)proton. From the scattering
azimuthal angle distribution, $\left<\mathcal{P}_yA_N\right>_{\rm scat}$ is found
to be $0.243 \pm 0.059 \pm 0.039$. The first uncertainty is
statistical, and the second is systematic. It is in good agreement
with the expected $\left<\mathcal{P}_yA_N\right>_{\rm exp} = 0.22 \pm 0.01$. This
validates the ability of a general-purpose spectrometer to function as
a proton polarimeter.

Besides BESIII, this approach might be used in the other
experiments. In the future facilities, such as
EIC~\cite{AbdulKhalek:2021gbh}, EicC~\cite{Anderle:2021wcy},
CEPC~\cite{CEPC}, and STCF~\cite{Achasov:2023gey} etc., this technique
should be considered in the early detector design stage, to optimize
the scattering layer for a higher efficiency of polarization
measurement without degrading the conventional detector function. The
successful integration of nucleon polarimeter function in a
general-purpose spectrometer enables the determination of spin
polarization of the final-state nucleon, and would expand the physics
capabilities of particle and nuclear physics experiments.

\begin{acknowledgments}
 
\textbf{Acknowledgement}

The BESIII Collaboration thanks the staff of BEPCII (https://cstr.cn/31109.02.BEPC) and the IHEP computing center for their strong support. This work is supported in part by National Key R\&D Program of China under Contracts Nos. 2023YFA1606800, 2023YFA1606704, 2025YFA1613900, 2023YFA1606000; National Natural Science Foundation of China (NSFC) under Contracts Nos. 12275320, 11975278, 11635010, 11935015, 11935016, 11935018, 12025502, 12035009, 12035013, 12061131003, 12192260, 12192261, 12192262, 12192263, 12192264, 12192265, 12221005, 12225509, 12235017, 12342502, 12361141819; the Chinese Academy of Sciences (CAS) Large-Scale Scientific Facility Program; the Strategic Priority Research Program of Chinese Academy of Sciences under Contract No. XDA0480600; CAS under Contract No. YSBR-101; 100 Talents Program of CAS; The Institute of Nuclear and Particle Physics (INPAC) and Shanghai Key Laboratory for Particle Physics and Cosmology; ERC under Contract No. 758462; Beijing Municipal Natural
Science Foundation under Contracts No. JQ22002; German Research Foundation DFG under Contract No. FOR5327; Istituto Nazionale di Fisica Nucleare, Italy; Knut and Alice Wallenberg Foundation under Contracts Nos. 2021.0174, 2021.0299, 2023.0315; Ministry of Development of Turkey under Contract No. DPT2006K-120470; National Research Foundation of Korea under Contract No. NRF-2022R1A2C1092335; National Science and Technology fund of Mongolia; Polish National Science Centre under Contract No. 2024/53/B/ST2/00975; STFC (United Kingdom); Swedish Research Council under Contract No. 2019.04595; U. S. Department of Energy under Contract No. DE-FG02-05ER41374.



\end{acknowledgments}

\bibliography{polarimeter}

@article{NakagawaAIP,
author = {Nakagawa,I. and others},
title = "{Polarization Measurements of RHIC‐pp RUN05 Using CNI pC‐Polarimeter}",
journal = {AIP Conference Proceedings},
volume = {915},
number = {1},
pages = {912-915},
year = {2007},
doi = {10.1063/1.2750924}
}

@article{PhysRevD.79.094014,
  title = "{Measurements of single and double spin asymmetry in $pp$ elastic scattering in the CNI region with a polarized atomic hydrogen gas jet target}",
  author = {Alekseev, I. G. and others},
  journal = {Phys. Rev. D},
  volume = {79},
  issue = {9},
  pages = {094014},
  numpages = {18},
  year = {2009},
  month = {May},
  publisher = {American Physical Society},
  doi = {10.1103/PhysRevD.79.094014},
  url = {https://link.aps.org/doi/10.1103/PhysRevD.79.094014}
}

@article{ComptonHERA,
	Author = {M Beckmann and others},
	Doi = {https://doi.org/10.1016/S0168-9002(01)00901-9},
	Issn = {0168-9002},
	Journal = {Nucl. Instr. Meth. A},
	Number = {2},
	Pages = {334-348},
	Title = "{The Longitudinal Polarimeter at HERA}",
	Url = {https://www.sciencedirect.com/science/article/pii/S0168900201009019},
	Volume = {479},
	Year = {2002}}

@article{ComptonJLab,
  title = "{Ultrahigh-precision Compton polarimetry at 2 GeV}",
  author = {Zec, A. and others},
  journal = {Phys. Rev. C},
  volume = {109},
  issue = {2},
  pages = {024323},
  numpages = {10},
  year = {2024},
  month = {Feb},
  publisher = {American Physical Society},
  doi = {10.1103/PhysRevC.109.024323},
  url = {https://link.aps.org/doi/10.1103/PhysRevC.109.024323}
}

@article{MottJLab2020,
  title = "{High precision 5 MeV Mott polarimeter}",
  author = {Grames, J. M. and others},
  journal = {Phys. Rev. C},
  volume = {102},
  issue = {1},
  pages = {015501},
  numpages = {24},
  year = {2020},
  month = {Jul},
  publisher = {American Physical Society},
  doi = {10.1103/PhysRevC.102.015501},
  url = {https://link.aps.org/doi/10.1103/PhysRevC.102.015501}
}

@article{Achasov:2023gey,
    author = "Achasov, M. and others",
    title = "{STCF conceptual design report (Volume 1): Physics \& detector}",    
    doi = "10.1007/s11467-023-1333-z",
    journal = "Front. Phys. (Beijing)",
    volume = "19",
    number = "1",
    pages = "14701",
    year = "2024"
}

@article{ABLIKIM2010345,
title = {Design and construction of the {BESIII} detector},
journal = {Nucl. Instr. Meth. A},
volume = {614},
number = {3},
pages = {345-399},
year = {2010},
issn = {0168-9002},
doi = {https://doi.org/10.1016/j.nima.2009.12.050},
url = {https://www.sciencedirect.com/science/article/pii/S0168900209023870},
author = {M. Ablikim and others}
}

@article{CAO2020163053,
title = "{Design and construction of the new BESIII endcap Time-of-Flight system with MRPC Technology}",
journal = {Nucl. Instr. Meth. A},
volume = {953},
pages = {163053},
year = {2020},
issn = {0168-9002},
doi = {https://doi.org/10.1016/j.nima.2019.163053},
url = {https://www.sciencedirect.com/science/article/pii/S0168900219314068},
author = {P. Cao and others}
}

@article{GROSNICK1990269,
title = {The design and performance of the FNAL high-energy polarized-beam facility},
journal = {Nucl. Instr. Meth. A},
volume = {290},
number = {2},
pages = {269-292},
year = {1990},
issn = {0168-9002},
doi = {https://doi.org/10.1016/0168-9002(90)90541-D},
url = {https://www.sciencedirect.com/science/article/pii/016890029090541D},
author = {D.P. Grosnick and others}
}

@article{PhysRevLett.90.142301,
  title = {Measurement of Spin-Correlation Parameters ${A}_{NN}$, ${A}_{SS}$, and ${A}_{SL}$ at 2.1 GeV in Proton-Proton Elastic Scattering},
  author = {Bauer, F. and others},
  collaboration = {EDDA Collaboration},
  journal = {Phys. Rev. Lett.},
  volume = {90},
  issue = {14},
  pages = {142301},
  numpages = {4},
  year = {2003},
  month = {Apr},
  publisher = {American Physical Society},
  doi = {10.1103/PhysRevLett.90.142301},
  url = {https://link.aps.org/doi/10.1103/PhysRevLett.90.142301}
}

@article{Bystricky:1976jr,
    author = "Bystricky, J. and Lehar, F. and Winternitz, P.",
    title = "{Formalism of Nucleon-Nucleon Elastic Scattering Experiments}",
    reportNumber = "SACLAY-DPHPE-76-12",
    doi = "10.1051/jphys:019780039010100",
    journal = "J. Phys. (France)",
    volume = "39",
    pages = "1",
    year = "1978"
}

@article{Anderle:2021wcy,
    author = "Anderle, Daniele P. and others",
    title = "{Electron-ion collider in China}",    
    reportNumber = "Frontiers of Physics, Volume 16 Issue (6):64701, 2021",
    doi = "10.1007/s11467-021-1062-0",
    journal = "Front. Phys. (Beijing)",
    volume = "16",
    number = "6",
    pages = "64701",
    year = "2021"
}

@article{AbdulKhalek:2021gbh,
    author = "Abdul Khalek, R. and others",
    title = "{Science Requirements and Detector Concepts for the Electron-Ion Collider}: {EIC Yellow Report}",
    primaryClass = "physics.ins-det",
    reportNumber = "BNL-220990-2021-FORE, JLAB-PHY-21-3198, LA-UR-21-20953",
    doi = "10.1016/j.nuclphysa.2022.122447",
    journal = "Nucl. Phys. A",
    volume = "1026",
    pages = "122447",
    year = "2022"
}

@article{KUNNE1988557,
title = "{Asymmetry in $\bar{p} p$ Elastic Scattering}",
journal = {Phys. Lett. B},
volume = {206},
number = {3},
pages = {557-560},
year = {1988},
issn = {0370-2693},
doi = {https://doi.org/10.1016/0370-2693(88)91629-2},
url = {https://www.sciencedirect.com/science/article/pii/0370269388916292},
author = {R.A. Kunne and others}
}

@article{PhysRevD.110.030001,
  title = {Review of Particle Physics},
  author = {Navas, S. and others},
  collaboration = {Particle Data Group Collaboration},
  journal = {Phys. Rev. D},
  volume = {110},
  issue = {3},
  pages = {030001},
  numpages = {5},
  year = {2024},
  month = {Aug},
  publisher = {American Physical Society},
  doi = {10.1103/PhysRevD.110.030001},
  url = {https://link.aps.org/doi/10.1103/PhysRevD.110.030001}
}

@article{PhysRev.148.1289,
  title = "{Polarization Parameter in $p\ensuremath{-}p$ Scattering from 328 to 736 MeV}",
  author = {Betz, F. and others},
  journal = {Phys. Rev.},
  volume = {148},
  issue = {4},
  pages = {1289--1296},
  numpages = {0},
  year = {1966},
  month = {Aug},
  publisher = {American Physical Society},
  doi = {10.1103/PhysRev.148.1289},
  url = {https://link.aps.org/doi/10.1103/PhysRev.148.1289}
}

@article{ALBROW1970445,
title = "{Polarization in elastic proton-proton scattering between 0.86 and 2.74 GeV/$c$}",
journal = {Nucl. Phys. B},
volume = {23},
number = {3},
pages = {445-465},
year = {1970},
issn = {0550-3213},
doi = {https://doi.org/10.1016/0550-3213(70)90296-8},
url = {https://www.sciencedirect.com/science/article/pii/0550321370902968},
author = {M.G. Albrow and others}
}

@article{PhysRevD.40.35,
  title = "{Elastic ${p}_{\ensuremath{\uparrow}}{p}_{\ensuremath{\uparrow}}$ scattering between 240 and 470 MeV}",
  author = {Onel, Y. and others},
  journal = {Phys. Rev. D},
  volume = {40},
  issue = {1},
  pages = {35--43},
  numpages = {0},
  year = {1989},
  month = {Jul},
  publisher = {American Physical Society},
  doi = {10.1103/PhysRevD.40.35},
  url = {https://link.aps.org/doi/10.1103/PhysRevD.40.35}
}

@article{PhysRevC.24.1778,
  title = "{$\mathrm{pp}$ elastic analyzing power from 318 to 800 MeV}",
  author = {McNaughton, M. W. and Chamberlin, E. P.},
  journal = {Phys. Rev. C},
  volume = {24},
  issue = {4},
  pages = {1778--1781},
  numpages = {0},
  year = {1981},
  month = {Oct},
  publisher = {American Physical Society},
  doi = {10.1103/PhysRevC.24.1778},
  url = {https://link.aps.org/doi/10.1103/PhysRevC.24.1778}
}

@article{vonPrzewoski:1998ye,
    author = "Przewoski, B. v.  and others",
    title = "{Proton proton analyzing power and spin correlation measurements between 250~MeV and 450~MeV at $7^\circ < \theta_{c.m.}< 90^\circ$  with an internal target in a storage ring}",
    doi = "10.1103/PhysRevC.58.1897",
    journal = "Phys. Rev. C",
    volume = "58",
    pages = "1897--1912",
    year = "1998"
}

@article{BESIII:2021cxx,
    author = "Ablikim, M. and others",
    collaboration = "BESIII Collaboration",
    title = "{Number of $J/\psi$ events at BESIII}",
    primaryClass = "hep-ex",
    doi = "10.1088/1674-1137/ac5c2e",
    journal = "Chin. Phys. C",
    volume = "46",
    number = "7",
    pages = "074001",
    year = "2022"
}

@article{PhysRevLett.41.384,
  title = "{Polarization Analyzing Power ${A}_{y}(\ensuremath{\theta})$ in $\mathrm{pp}$ Elastic Scattering at 643, 787, and 796 MeV}",
  author = {Bevington, P. R. and others},
  journal = {Phys. Rev. Lett.},
  volume = {41},
  issue = {6},
  pages = {384--387},
  numpages = {0},
  year = {1978},
  month = {Aug},
  publisher = {American Physical Society},
  doi = {10.1103/PhysRevLett.41.384},
  url = {https://link.aps.org/doi/10.1103/PhysRevLett.41.384}
}

@article{PhysRev.105.288,
  title = "{Experiments with 315-MeV Po\-lari\-zed Protons: Proton-Proton and Proton-Neutron Scattering}",
  author = {Chamberlain, O. and others},
  journal = {Phys. Rev.},
  volume = {105},
  issue = {1},
  pages = {288--301},
  numpages = {0},
  year = {1957},
  month = {Jan},
  publisher = {American Physical Society},
  doi = {10.1103/PhysRev.105.288},
  url = {https://link.aps.org/doi/10.1103/PhysRev.105.288}
}

@article{GREENIAUS1979308,
title = "{Measurements of p-p and p-$^4$He ana\-ly\-zing powers at medium energies}",
journal = {Nucl. Phys. A},
volume = {322},
number = {2},
pages = {308-328},
year = {1979},
issn = {0375-9474},
doi = {https://doi.org/10.1016/0375-9474(79)90428-7},
url = {https://www.sciencedirect.com/science/article/pii/0375947479904287},
author = {L.G. Greeniaus and others},
}

@article{PhysRev.163.1470,
  title = "{Nucleon-Nucleon Polarization between 300 and 700 MeV}",
  author = {Cheng, David and others},
  journal = {Phys. Rev.},
  volume = {163},
  issue = {5},
  pages = {1470--1478},
  numpages = {0},
  year = {1967},
  month = {Nov},
  publisher = {American Physical Society},
  doi = {10.1103/PhysRev.163.1470},
  url = {https://link.aps.org/doi/10.1103/PhysRev.163.1470}
}

@article{PhysRev.95.1348,
  title = "{Small-Angle $p\ensuremath{-}p$ Cross Sections and Polarization at 300 MeV}",
  author = {Chamberlain, O. and others},
  journal = {Phys. Rev.},
  volume = {95},
  issue = {5},
  pages = {1348--1349},
  numpages = {0},
  year = {1954},
  month = {Sep},
  publisher = {American Physical Society},
  doi = {10.1103/PhysRev.95.1348},
  url = {https://link.aps.org/doi/10.1103/PhysRev.95.1348}
}

@article{PhysRevD.21.580,
  title = "{Measurement of the spin-dependent parameters $D$, $R$, $A$, and $P$ for small-angle $p\ensuremath{-}p$ elastic scattering between 300 and 600 MeV}",
  author = {Besset, D. and others},
  journal = {Phys. Rev. D},
  volume = {21},
  issue = {3},
  pages = {580--598},
  numpages = {0},
  year = {1980},
  month = {Feb},
  publisher = {American Physical Society},
  doi = {10.1103/PhysRevD.21.580},
  url = {https://link.aps.org/doi/10.1103/PhysRevD.21.580}
}

@article{PhysRevC.76.025209,
  title = {Updated analysis of $NN$ elastic scattering to 3 GeV},
  author = {Arndt, R. A. and others},
  journal = {Phys. Rev. C},
  volume = {76},
  issue = {2},
  pages = {025209},
  numpages = {10},
  year = {2007},
  month = {Aug},
  publisher = {American Physical Society},
  doi = {10.1103/PhysRevC.76.025209},
  url = {https://link.aps.org/doi/10.1103/PhysRevC.76.025209}
}

@misc{GWDAC,
  title = "{SAID database}",
  howpublished = "\url{https://gwdac.phys.gwu.edu}",
}

@article{PhysRevLett.129.131801,
  title = "{Precise Measurements of Decay Parameters and $CP$ Asymmetry with Entangled $\mathrm{\ensuremath{\Lambda}}\text{\ensuremath{-}}\overline{\mathrm{\ensuremath{\Lambda}}}$ Pairs}",
  author = {Ablikim, M. and others},
  collaboration = {BESIII Collaboration},
  journal = {Phys. Rev. Lett.},
  volume = {129},
  issue = {13},
  pages = {131801},
  numpages = {8},
  year = {2022},
  month = {Sep},
  publisher = {American Physical Society},
  doi = {10.1103/PhysRevLett.129.131801},
  url = {https://link.aps.org/doi/10.1103/PhysRevLett.129.131801}
}

@article{PhysRevLett.127.012003,
  title = "{Cornucopia of Antineutrons and Hyperons from a Super $J/\ensuremath{\psi}$ Factory for Next-Generation Nuclear and Particle Physics High-Precision Experiments}",
  author = {Yuan, Chang-Zheng and Karliner, Marek},
  journal = {Phys. Rev. Lett.},
  volume = {127},
  issue = {1},
  pages = {012003},
  numpages = {6},
  year = {2021},
  month = {Jun},
  publisher = {American Physical Society},
  doi = {10.1103/PhysRevLett.127.012003},
  url = {https://link.aps.org/doi/10.1103/PhysRevLett.127.012003}
}

@article{BESIII:2024geh,
    author = "Ablikim, Medina and others",
    collaboration = "BESIII Collaboration",
    title = "{First Study of Antihyperon-Nucleon Scattering \ensuremath{\bar\Lambda}p\textrightarrow{}\ensuremath{\bar\Lambda}p and Measurement of \ensuremath{\Lambda}p\textrightarrow{}\ensuremath{\Lambda}p Cross Section}",
    primaryClass = "hep-ex",
    doi = "10.1103/PhysRevLett.132.231902",
    journal = "Phys. Rev. Lett.",
    volume = "132",
    number = "23",
    pages = "231902",
    year = "2024"
}

@article{BESIII:2023clq,
    author = "Ablikim, Medina and others",
    collaboration = "BESIII Collaboration",
    title = "{First Study of Reaction \ensuremath{\Xi}$^{0}$n\textrightarrow{}\ensuremath{\Xi}$^{-}$p Using \ensuremath{\Xi}$^{0}$-Nucleus Scattering at an Electron-Positron Collider}",
    primaryClass = "hep-ex",
    doi = "10.1103/PhysRevLett.130.251902",
    journal = "Phys. Rev. Lett.",
    volume = "130",
    number = "25",
    pages = "251902",
    year = "2023"
}

@article{ONUKI202278,
title = "{Belle-II status and prospect}",
journal = {Nucl. Part. Phys. Proc.},
volume = {318-323},
pages = {78-84},
year = {2022},
issn = {2405-6014},
doi = {https://doi.org/10.1016/j.nuclphysbps.2022.09.017},
url = {https://www.sciencedirect.com/science/article/pii/S2405601422000177},
author = {Yoshiyuki Onuki}
}

@article{Hayrapetyan_2024,
doi = {10.1088/1748-0221/19/05/P05064},
url = {https://dx.doi.org/10.1088/1748-0221/19/05/P05064},
year = {2024},
month = {may},
publisher = {IOP Publishing},
volume = {19},
number = {05},
pages = {P05064},
author = {Hayrapetyan, A. and others},
collaboration = {CMS Collaboration},
title = "{Development of the CMS detector for the CERN LHC Run 3}",
journal = {JINST}
}

@article{Aad_2024,
doi = {10.1088/1748-0221/19/05/P05063},
url = {https://dx.doi.org/10.1088/1748-0221/19/05/P05063},
year = {2024},
month = {may},
publisher = {IOP Publishing},
volume = {19},
number = {05},
pages = {P05063},
author = {Aad, G. and others},
collaboration = {ATLAS Collaboration},
title = "{The ATLAS experiment at the CERN Large Hadron Collider: a description of the detector configuration for Run 3}",
journal = {JINST}
}

@article{doi:10.1142/S0217751X15300227,
author = {R. Aaij and others},
collaboration = {LHCb Collaboration},
title = "{LHCb detector performance}",
journal = {Int. J Mod. Phys. A},
volume = {30},
number = {07},
pages = {1530022},
year = {2015},
doi = {10.1142/S0217751X15300227},

URL = {https://doi.org/10.1142/S0217751X15300227 }
}

@article{c642-1lzb,
  title = {How to determine nucleon polarization at existing collider experiments?},
  author = {Liang, Yu-Tie and others},
  journal = {Phys. Rev. D},
  volume = {112},
  issue = {3},
  pages = {L031502},
  numpages = {6},
  year = {2025},
  month = {Aug},
  publisher = {American Physical Society},
  doi = {10.1103/c642-1lzb},
  url = {https://link.aps.org/doi/10.1103/c642-1lzb}
}

@article{CEPC,
    author = "Dong, Mingyi and others",
    title = "{CEPC Conceptual Design Report}",
    eprint = "1811.10545",
    archivePrefix = "arXiv",
    primaryClass = "hep-ex",
    month = "11",
    year = "2018",
    journal = ""
}

@article{Chen:2024aom,
    author = "Chen, Jinhui and others",
    title = "{Properties of the QCD matter: review of selected results from the relativistic heavy ion collider beam energy scan (RHIC BES) program}",
    primaryClass = "nucl-ex",
    doi = "10.1007/s41365-024-01591-2",
    journal = "Nucl. Sci. Tech.",
    volume = "35",
    number = "12",
    pages = "214",
    year = "2024"
}

@article{PhysRevLett.88.092301,
  title = "{Measurement of ${G}_{{E}_{p}}/{G}_{{M}_{p}}$ in $\stackrel{\ensuremath{\rightarrow}}{e}\mathit{p}\ensuremath{\rightarrow}\mathit{e}\stackrel{\ensuremath{\rightarrow}}{p}$ to ${\mathit{Q}}^{2}\phantom{\rule{0ex}{0ex}}=\phantom{\rule{0ex}{0ex}}5.6~{\mathrm{GeV}}^{2}$}",
  author = {Gayou, O. and others},
  collaboration = {Jefferson Lab Hall A Collaboration},
  journal = {Phys. Rev. Lett.},
  volume = {88},
  issue = {9},
  pages = {092301},
  numpages = {5},
  year = {2002},
  month = {Feb},
  publisher = {American Physical Society},
  doi = {10.1103/PhysRevLett.88.092301},
  url = {https://link.aps.org/doi/10.1103/PhysRevLett.88.092301}
}

@misc{iminuit,
  author={Hans Dembinski and Piti Ongmongkolkul et al.},
  title={scikit-hep/iminuit},
  DOI={10.5281/zenodo.3949207},
  publisher={Zenodo},
  year={2020},
  month={Dec},
  url={https://doi.org/10.5281/zenodo.3949207}
}

@article{James:1975dr,
    author = "James, F. and Roos, M.",
    title = "{Minuit: A System for Function Minimization and Analysis of the Parameter Errors and Correlations}",
    reportNumber = "CERN-DD-75-20",
    doi = "10.1016/0010-4655(75)90039-9",
    journal = "Comput. Phys. Commun.",
    volume = "10",
    pages = "343--367",
    year = "1975"
}

@article{PhysRevC.62.034005,
  title = "{Nucleon-nucleon elastic scattering to 3 GeV}",
  author = {Arndt, Richard A. and Strakovsky, Igor I. and Workman, Ron L.},
  journal = {Phys. Rev. C},
  volume = {62},
  issue = {3},
  pages = {034005},
  numpages = {16},
  year = {2000},
  month = {Aug},
  publisher = {American Physical Society},
  doi = {10.1103/PhysRevC.62.034005},
  url = {https://link.aps.org/doi/10.1103/PhysRevC.62.034005}
}

@article{GlueX:2020idb,
    author = "Adhikari, S. and others",
    collaboration = "GlueX",
    title = "{The GLUEX beamline and detector}",    
    primaryClass = "physics.ins-det",
    reportNumber = "JLAB-PHY-20-3195",
    doi = "10.1016/j.nima.2020.164807",
    journal = "Nucl. Instrum. Meth. A",
    volume = "987",
    pages = "164807",
    year = "2021"
}

@article{Dai:2024myk,
    author = "Dai, Jianping and others",
    title = "{Prospects to study hyperon-nucleon interactions at BESIII}",
    eprint = "2209.12601",
    archivePrefix = "arXiv",
    primaryClass = "hep-ex",
    doi = "10.1088/1674-1137/ad3dde",
    journal = "Chin. Phys. C",
    volume = "48",
    number = "7",
    pages = "073003",
    year = "2024"
}

@article{LANGE2001152,
title = "{The EvtGen particle decay simulation package}",
journal = {Nucl. Instr. Meth. A},
volume = {462},
number = {1},
pages = {152-155},
year = {2001},
issn = {0168-9002},
doi = {https://doi.org/10.1016/S0168-9002(01)00089-4},
url = {https://www.sciencedirect.com/science/article/pii/S0168900201000894},
author = {David J. Lange}
}

@article{PingRong-Gang_2008,
doi = {10.1088/1674-1137/32/8/001},
url = {https://dx.doi.org/10.1088/1674-1137/32/8/001},
year = {2008},
month = {aug},
publisher = {},
volume = {32},
number = {8},
pages = {599},
author = {Rong-Gang Ping},
title = "{Event generators at BESIII}",
journal = {Chin. Phy. C}
}

@article{PhysRevD.87.012002,
  title = {Search for hadronic transition ${\ensuremath{\chi}}_{cJ}\ensuremath{\rightarrow}{\ensuremath{\eta}}_{c}{\ensuremath{\pi}}^{+}{\ensuremath{\pi}}^{\ensuremath{-}}$ and observation of ${\ensuremath{\chi}}_{cJ}\ensuremath{\rightarrow}K\overline{K}\ensuremath{\pi}\ensuremath{\pi}\ensuremath{\pi}$},
  author = {Ablikim, M. and others},
  collaboration = {BESIII Collaboration},
  journal = {Phys. Rev. D},
  volume = {87},
  issue = {1},
  pages = {012002},
  numpages = {13},
  year = {2013},
  month = {Jan},
  publisher = {American Physical Society},
  doi = {10.1103/PhysRevD.87.012002},
  url = {https://link.aps.org/doi/10.1103/PhysRevD.87.012002}
}

@article{AGOSTINELLI2003250,
title = {Geant4—a simulation toolkit},
journal = {Nucl. Instr. Meth. A},
volume = {506},
number = {3},
pages = {250-303},
year = {2003},
issn = {0168-9002},
doi = {https://doi.org/10.1016/S0168-9002(03)01368-8},
url = {https://www.sciencedirect.com/science/article/pii/S0168900203013688},
author = {S. Agostinelli and others}
}

\end{document}